\documentclass[12pt,preprint]{aastex}
\shortauthors{Hsia et al.}
\shorttitle{Binary origin of the Young Planetary Nebula HB 12}

\received{}
\begin{document}

\title{Evidence for a Binary origin of the Young Planetary Nebula HB 12}
\author{Chih Hao Hsia$^{1,2}$,Wing Huen Ip$^{1}$,Jin Zeng Li$^{2}$}
\affil{$^{1}$Institute of Astronomy, National Central University, Chung Li
             32054, Taiwan (E-mail: d929001@astro.ncu.edu.tw) \\
$^{2}$National Astronomical Observatories, Chinese Academy of
           Sciences, Beijing 100012, China (E-mail: ljz@bao.ac.cn) \\}

\begin{abstract}
Young planetary nebulae play an important role in stellar
evolution when intermediate- to low-mass stars (0.8 $\sim$ 8
M$_\odot$) evolve from the proto-planetary nebulae phase to the
planetary nebulae phase. Many young planetary nebulae display
distinct bipolar structures as they evolve away from the
proto-planetary nebulae phase. One possible cause of their
bipolarity could be due to a binary origin of its energy source.
Here we report our detailed investigation of the young planetary
nebula, Hubble 12, which is well-known for its extended
hourglass-like envelope.  We present evidence with time-series
photometric observations the existence of an eclipsing binary at
the center of Hubble 12.  Low-resolution spectra of the central
source show, on the other hand, absorption features such as CN,
G-band \& Mg b${\arcsec}$, which can be suggestive of a low-mass
nature of the secondary component.

\end{abstract}


\section{Introduction}

The young planetary nebula (PN) Hubble 12 (HB 12; PN G111.8-02.8) plays an important
role in the study of PNe. The line ratios of its strong fluorescent molecular hydrogen
emission (Dinerstein et al., 1988) match closely pure fluorescent emission (Black and
van Dishoeck, 1987), which might have originated from shock excitation via collisional
interaction of a strong wind from the central star against a circumstellar gaseous disk
(Kastner et al., 1994). Particular interest also comes from its bipolar structure of HB 12.
Early radio continuum observations from VLA (Bignell, 1983) first disclosed its bipolar
configuration.  Miranda and Solf (1989) later confirmed the double lobe structure along
the north-south axis through long-slit spectroscopy.
In addition, Hora \& Latter (1996) and Welch et al. (1999) showed the existence of a
ring-like structure near the central core from near-infrared and [\ion{Fe}{2}] imaging.
Recent HST/NICMOS observations have revealed clearly detailed structure of the inner
torus and its bipolar lobes (Hora et al., 2000).  The origin of the axis-symmetric
morphology of young PNe has long been an unresolved issue.  In classification of
PNe based on their morphology, Zuckerman and Aller (1986) and Soker (1997) found
that a large majority of PNe have non-spherical shapes, some indicate extreme bipolar
or axis-symmetric appearance. Such structures could be generated by strong bipolar outflows
during the late AGB or post-AGB phase and may be a transient phenomineon (Kwok et al., 2000).
The bipolar flows in turn could be produced by close binaries (Morris, 1990; Soker et al., 1998;
Soker, 2000) or rather focusing effects from the associated magnetic field (Garcia-Segura et al.,
1999, 2000) among various potential mechanisms suggested. Note that observational evidence
has been found in support of the binary model (De Marco et al., 2004; Hillwig, 2004; Sorensen
and Pollacco, 2004).  Detailed investigations of the physical nature and dynamical properties
of the central source is therefore of fundamental importance to our understanding of the
origin and evolution of PNe.

We have initiated a program combining efforts from the photometric
measurements obtained by the one-meter telescope (LOT) on the
Lulin Observatory at central Taiwan and the spectrographic
observations carried out with the 2.16 m telescope at the
Xing-Long station of the National Astronomical Observatory of the
Chinese Academy of Sciences (NAOC). Our main objective is to
search for the possible presence of periodic variations in the
lightcurves of the nucleus of HB 12, NSV 26083. Properties of the
binary components could also be inferred from perhaps spectral
observations of the central source.  We describe, in Section 2,
the observations and data reduction. The photometric lightcurves
are presented in Section 3, and then a discussion of the results
and interpretations in Section 4.  The spectral features possibly
originating from the cool secondary are investigated in Section 5,
which is followed by a summary of the main results of this study.

\section{Observations and Data Reduction}

\subsection{Time-series Broadband Photometric Imaging}

High-speed broadband photometric observations were performed in
the queue mode on the nights of December 3-5, 2003 using Johnson R
and I band filters with the LOT telescope of the National Central
University at Taiwan. The camera was operated with a Princeton
Instruments 1340 $\times$ 1300 pixel CCD, giving a field of view
of 11$\arcmin \times 11\arcmin$. The CCD has a readout noise of
15.7e and a gain of 4.4. The setup results in a pixel scale of
0.62$\arcsec$ pixel$^{-1}$. Flat field exposures were obtained on
the twilight sky. The seeing condition all through this run of
observations varied between 1.3$\arcsec$ and 1.9$\arcsec$. The
journal of observations is summarized in Table 1.

More than 800 snap shots of HB 12 were made, with each snap shot
containing a 10s exposure in R and another 5s in I. The data
reduction includes bias and dark current correction and
flat-fielding based on standard packages and procedures in the
NOAO IRAF (V2.12).  Differential magnitudes of the central core of
HB 12 were measured using the DAOPHOT package with three stars in
the same field as reference stars. The S/N ratios of all stars are
$>$ 100.  No apparent variations were found with the reference
stars. The differential magnitudes in the R \& I bands have an
accuracy within 0.04 mag and 0.02 mag, respectively.

\subsection{Optical Spectroscopy}

Low-resolution spectroscopy was obtained by the 2.16 m telescope
of NAOC. The journal of observations in two separate sessions is given
in Table 2.  In the 2004 session, measurement was performed in the spectral
range (4800 - 10500 \AA) on the night of Aug. 8, 2004.
The spectral dispersion was 3.1 \AA~pixel$^{-1}$.
A Beijing Faint Object Spectrograph and Camera
(BFOSC) and a thinned back-illuminated Orbit 2048 $\times$ 2048 CCD were used.
A slit width of 3.6$\arcsec$ was set. The exposure times ranged from 300 to 900 s.
S/N ratios of the continuum of $>$ 90 were achieved. Exposures of Fe-Ne arcs
were obtained right before and after each stellar spectrum and used for the
wavelength calibration.

In the 2005 session, spectroscopy was performed in the blue
(3800 - 6200\AA) spectral range on the night of Sep. 26, 2005. The
100 \AA~mm$^{-1}$ grating was used, which result in a two-pixel resolution
of $\sim$ 4.8 \AA. The slit was placed at P.A. = 170$^\circ$, in parallel to
the main axis of HB 12 as indicated in Figure 1.
The slit width was set to 2$\arcsec$.  An Optomechanics Research Inc. spectrograph
and a Tektronix 1024 $\times$ 1024 CCD were used. The exposures ranged from
3600 to 7200 s, resulting in S/N ratios of $>$ 60.  Wavelength calibration was
performed based on He-Ar lamps exposed right before and after the target spectrum.

The spectral data were reduced following standard procedures in
the NOAO IRAF (V2.12) software package. The CCD reductions
included bias and flat-field correction, successful background
subtraction and cosmic-ray removal. Flux calibration was derived
with observations of at least two of the KPNO standard stars per
spectral range per night. The atmospheric extinction was corrected
by the mean extinction coefficients measured for Xing-Long
station, where the 2.16 m telescope is located.

\section{Search for periodicities in the lightcurves}

The multi-band photometric results of HB 12 were presented in Figure 2.
The method of phase dispersion minimization (PDM, Stellingwerf, 1978) was
used to analyze the lightcurves of NSV 26083. The PDM code was employed
to derive the period, maximum magnitude and amplitude of the light variation.
Before calculating the power spectra, we set the nightly mean magnitude to
zero and calculated the amplitude spectra of these data.  The I band data clearly
covers three primary minima.  A linear least square fit results in a period of
P = 0.1415 $\pm$ 0.0015 days. The R band lightcurve was fitted simultaneously
using the period of the I band data because of its distinct existence of primary minima.
The power spectra of the
photometric data were presented in Figure 3. A prominent amplitude peak is found
at 7.06 c/d, which is 3.4 hrs. The corresponding phase diagrams are shown in
Figure 4. The profiles are sinusoidal for both the R and I bands.  The period shows
an amplitude of 0.06 $\pm$ 0.0074 mag in R and 0.08 $\pm$ 0.0046 mag in I. This
suggests that NSV 26083 displays in its multi-band time-series observations periodic
variation and indicates probably an eclipsing binary origin of HB 12.  It is the
first time that a clear signature of periodicity is evidenced toward the exciting
source of HB 12, which may have important implications on the physical nature of
bipolar structures associated with other young PNe.

Note that the dip in the R band lightcurve seems to be shallower and smoother
than that in the I band lightcurve.  This wavelength dependence could probably be
due to effects from H$\alpha$ emission of the companion star, which is encompassed
by the R band observations, or otherwise reflection effects from the illuminated surface
of the less luminous star (Grauer \& Bond, 1983; Bruch et al, 2001).

If we alternatively suppose the periodic variations of NSV 26083 are due to rotation
modulation of stellar spot(s), the rotational period can be estimated as follows (Reid et al., 1993):

\begin{equation}
P_{crit} = \frac{2\pi R_{e}^{3/2}}{\sqrt{GM}} = 2.78 R_{e}^{3/2} M^{-1/2}
\end{equation}

where P$_{crit}$ is the rotational period of the star in unit of
hrs. R$_{e}$ is the equatorial radius and M the mass of the star
in solar units. Following the discussion by Reid et al. (1993), we
use R$_{e}$ = 1.5 R$_{*}$, where R$_{*}$ is the radius of NSV
26083.  If Zhang and Kwok's (1993) result for T$_{eff}$ = 31800 K,
$\log$ g = 3.1 and M = 0.8 M$_\odot$ are adopted, the rotational
period P$_{crit}$ would be as large as 48 hrs. This is
inconsistent with our estimation of 3.4 hrs as presented above and
helps to exclude the possibility of the modulation by stellar
spots.

\section{Stellar properties of the binary components}

The mass of the central source of HB 12 has been determined to be
0.8 M$_\odot$ by Zhang and Kwok (1993) based on existing infrared and radio
data. We present below an estimation of the mass and radius of the proposed
secondary component of the system.

First, the secondary star with an orbital period of hours to days was suggested
to have a mass of less than 0.5 M$_\odot$ (Chen et al., 1995). If a mass ratio
M$_{2}$ / M$_{1}$ $<$ 0.8 (M$_{1}$ is the mass of the primary star) is supposed,
the upper limits of the mass and radius of the secondary would satisfy the following condition
(Paczy$\acute{n}$ski, 1981):

\begin{equation}
8.85 \sqrt{\frac{R_{2}^{3}}{M_{2}}} < P
\end{equation}

where P is the orbital period in hrs. M$_{2}$ and R$_{2}$ are the mass and
radius of the secondary in solar units, respectively.

Second, assume that the mass-radius relation of the lower main sequence stars can be
applied to the secondary (Rappaport et al, 1982):

\begin{equation}
\frac{R_{2}}{R_\odot} = 0.76 (\frac{M_{2}}{M_\odot})^{0.78}
\end{equation}

We can combine Eq. (2) with Eq. (3) to obtain

\begin{equation}
M_{2} < 0.443 M_\odot~and~R_{2} < 0.403 R_\odot
\end{equation}

There is a general agreement (Sch$\ddot{o}$nberner, 1981; Heap \&
Augensen, 1987; Weidemann, 1989; Tylenda et al., 1991b; Zhang \&
Kwok, 1993; Stasi$\acute{n}$ska et al., 1997) that the dispersion
of the central stellar masses of planetary nebulae, averaged to
around 0.6 M$_\odot$, should be rather small. If we assume a mass
0.6 M$_\odot$ of the primary star, for M$_{1}$ = 0.6 and 0.08 $<$
M$_{2}$ $<$ 0.443 , the separation of the stars a = 1.163 $\pm$
0.063 R$_\odot$.  In turn, the radius of the Roche lobe of the
secondary is $\ell$$_{2}$ = a [ 0.5 + 0.227 $\log$ (M$_{2}$ /
M$_{1}$) ]. This results in an estimation of 0.547 $\pm$ 0.03
R$_\odot$ and the hemisphere of the secondary must be illuminated
and heated by the primary source.

The distance to HB 12 has been estimated to be 2.24 kpc by Cahn et
al. (1992) based on existing optical and radio data, Hora and
Latter (1996) determined a E($\bv$) value is 0.28 measuring from
the Brackett line flux of the near-IR spectrum, the V magnitude
determined by Tylenda et al. (1991a) for the cool stellar
component of HB 12 is 13.6. If these results are adopted, the
absolute visual magnitude M$_{v}$ of the central star of HB 12
will be 0.98.  Suppose that the primary component has a mass of
0.6 M$_\odot$ and M$_{v}$ = 0.98 at an effective temperature of
31800 K (Zhang and kwok, 1993), the corresponding radius of the
primary is estimated to be R$_{1}$ = 0.19 R$_\odot$.

\section{The Spectra of NSV 26083}

In order to examine the nature of the putative binary companion of
NSV 26083, we have initiated a project of spectrographic
measurements at the National Astronomical Observatory of the
Chinese Academy of Sciences using the 2.16 m telescope. Figure 5
shows the low-resolution spectrum of NSV 26083 taken on August 8,
2004, which apparently indicates various emission lines
characteristic of photoionized medium. The profiles of H$\alpha$
and H$\beta$ are broader than other emission lines, which is here
attributed to most likely effects of the Rayleigh-Raman scatting
(Arrieta $\&$ Torres-Peimbert, 2003).

To search for further evidence on the possible binary origin of
the nucleus of HB 12, we examine closely the spectra taken on
September 26, 2005 photospheric absorption features characteristic
of a cool companion. The spectrum between 4150 and 4550 \AA~is
shown in Figure 6a in an expanded scale. Apparent G-band feature
characteristic of late type stars is seen, and the molecular CN
$\lambda$4216 absorption can also be identified. Furthermore,
absorption features due to the s-process elements such as Y and Sr
were found in the spectrum with \ion{Y}{2} $\lambda$4178 and
\ion{Sr}{1} $\lambda$4607 being the primary features.  The
molecular C$^{13}$C$^{13}$ $\lambda$4752 is also seen in
absorption in the spectrum as shown in Figure 6b. The spectrum
ranges from 5150 to 5600 \AA~clearly indicates C$_{2}$ features at
$\lambda$5165 and $\lambda$5585 and is presented in Figure 6c. The
\ion{Mg}{1} triplet ($\lambda$$\lambda$5167-72-83) is also
marginally seen in the spectrum.

The above mentioned features seems to suggest the existence of a
cool companion with a spectral type of G to early K to the
exciting source of HB 12.  However, this introduces discrepancy
with our mass estimation based on the lightcurves, which gives a
spectral type of M.  Note that a M dwarf in isolation can not be
detected at all at the distance of HB 12 of about 2.24 kpc (Cahn
et al., 1992).  This discrepancy can not be reconciled unless
additional physical processes are involved.  The spectral change
of the cool secondary is here attributed to most likely external
heating of its upper atmosphere by the hot primary (Grauer and
Bond, 1983). Further investigations of this system based on high
resolution spectroscopic observations is highly needed to have
this issued resolved, which may come up with a more reliable
determination of the spectral type of the secondary. However, this
uncertainty with the spectral determination does not affect in any
way our inference of the binary origin of the nucleus of HB 12
based on our photometric results.

\section{Summary}

Based on the time-series multi-band photometric observations and
low-resolution spectroscopy, our study of the physical nature of
the nucleus (NSV 26083) of the young PN HB 12 has led to the
following results:

1. The central star is probably a close binary with an orbital
period of 3.4 hrs.  This provides further support to the theory of
a binary origin of bipolar PNe. \\

2. The difference in the R and I lightcurves is indicative of a reflection
effect of the illuminated surface of the secondary.\\

3. Assuming a mass of 0.6 M$_\odot$ of the primary, an upper limit
mass and radius of the secondary star can be estimated to be
M$_{2}$ $<$ 0.443 M$_\odot$ and R$_{2}$ $<$ 0.403 R$_\odot$,
respectively. This results in an estimation of a physical
separation of $\sim$ 1.163 R$_\odot$ of the close binary in
association with NSV 26083. The hemisphere of the secondary can be
suffering from reflection and heating effects from the hot
primary. This thermal coupling may well lead to a spectral change
of the secondary and deserve to be investigated further in detail.\\

{\flushleft \bf Acknowledgments~}

We are grateful to the Reviewer for useful comments, and Prof. Sun
Kwok at University of Hong Kong, Prof. Yi Chou at National Central
University, and Dr. Yu-Lei Qiu, Dr. Jian-Yan Wei, and Prof.
Jing-Yao Hu at National Astronomical Observatory of the Chinese
Academy of Sciences
 for useful discussions. This work was partially supported by the National Science
Council of Taiwan under NSC 93-2752-M-008-001-PAE, NSC
93-2112-M-008-006, NSC 94-2752-M-008-001-PAE, and NSC
94-2112-M-008-002. Finally, we acknowledge funding from the
National Natural Science Foundation of China through grant
O611081001.

\bibliographystyle{aa}

\clearpage

\figcaption[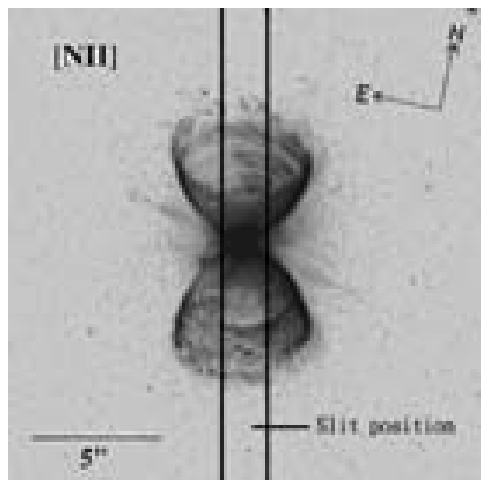]{The HST/WFPC2 narrow-band [\ion{N}{2}] (F658N) image of HB 12, displayed
with a linear gray scale. We present here the combined data sets
(U6CI0405, U6CI0406, U6CI0407,and U6CI0408) of B. Balick. The total exposure
time is 1300 s and the field of view is 18$\arcsec \times 18\arcsec$.  The slit
position is shown against the image of the core of the PN. \label{fig.1}}

\figcaption[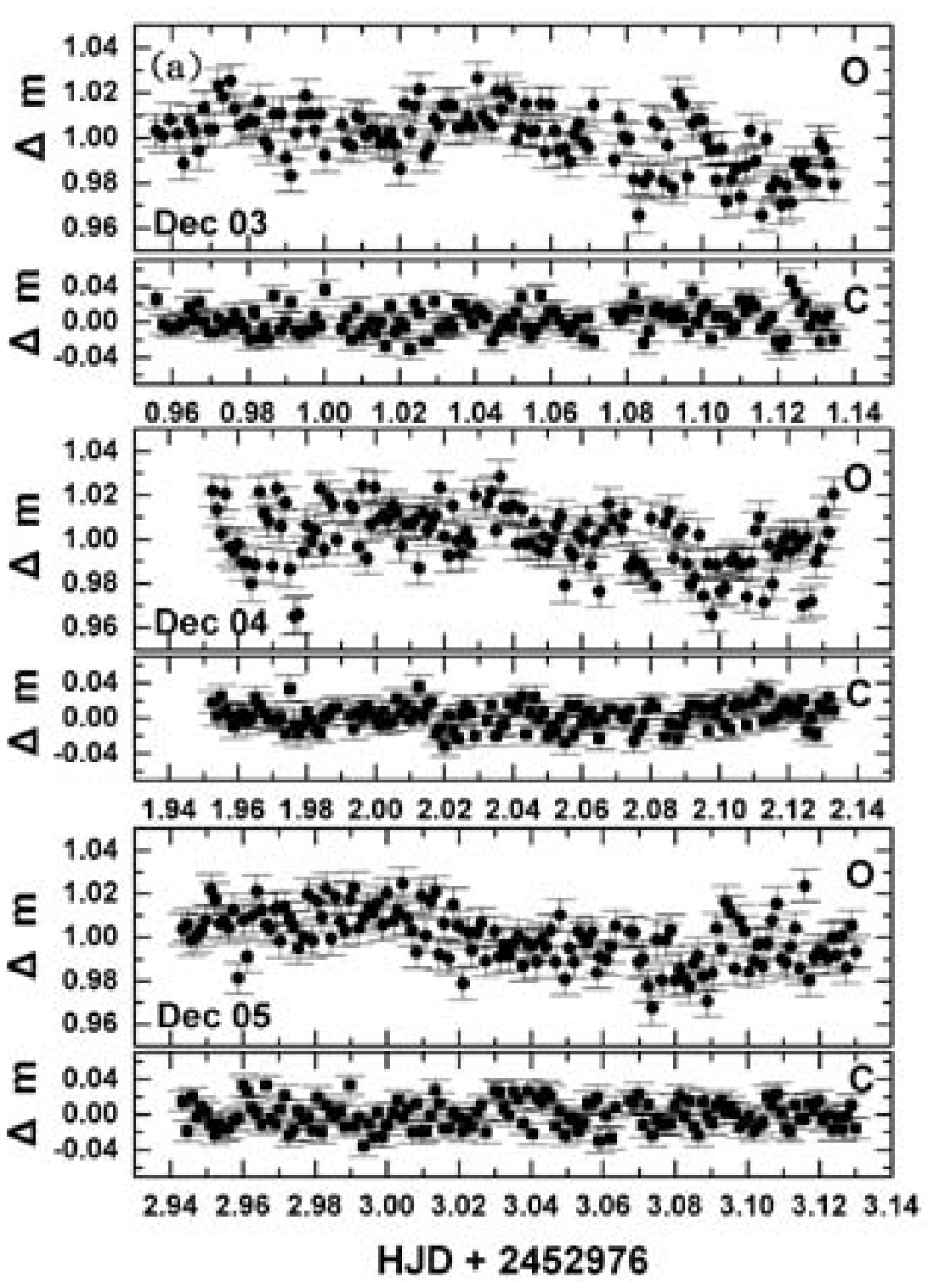,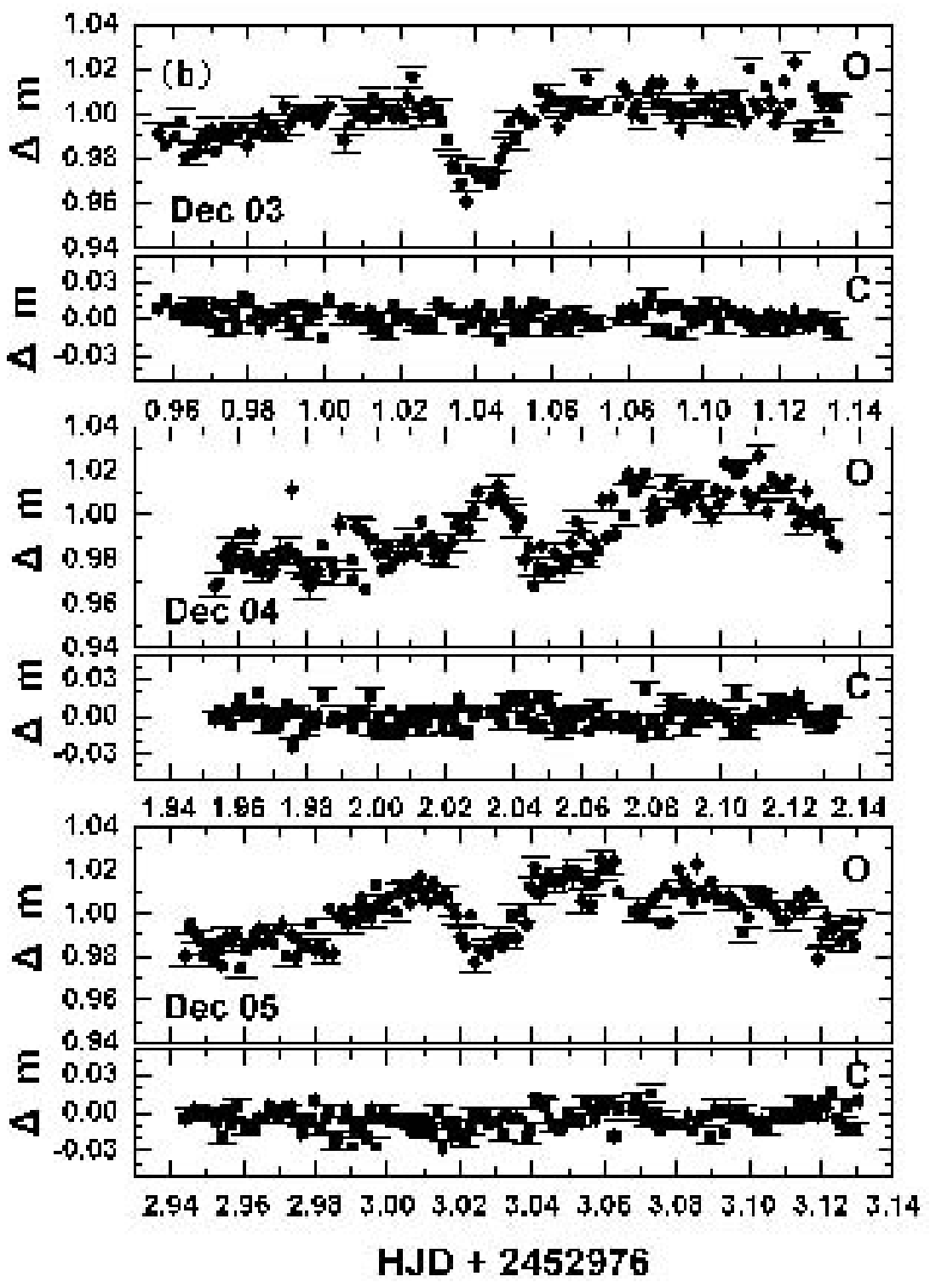]{Differential photometric lightcurves of NSV 26083 in both the R band (a)
and the I band (b). \label{fig.2}}

\figcaption[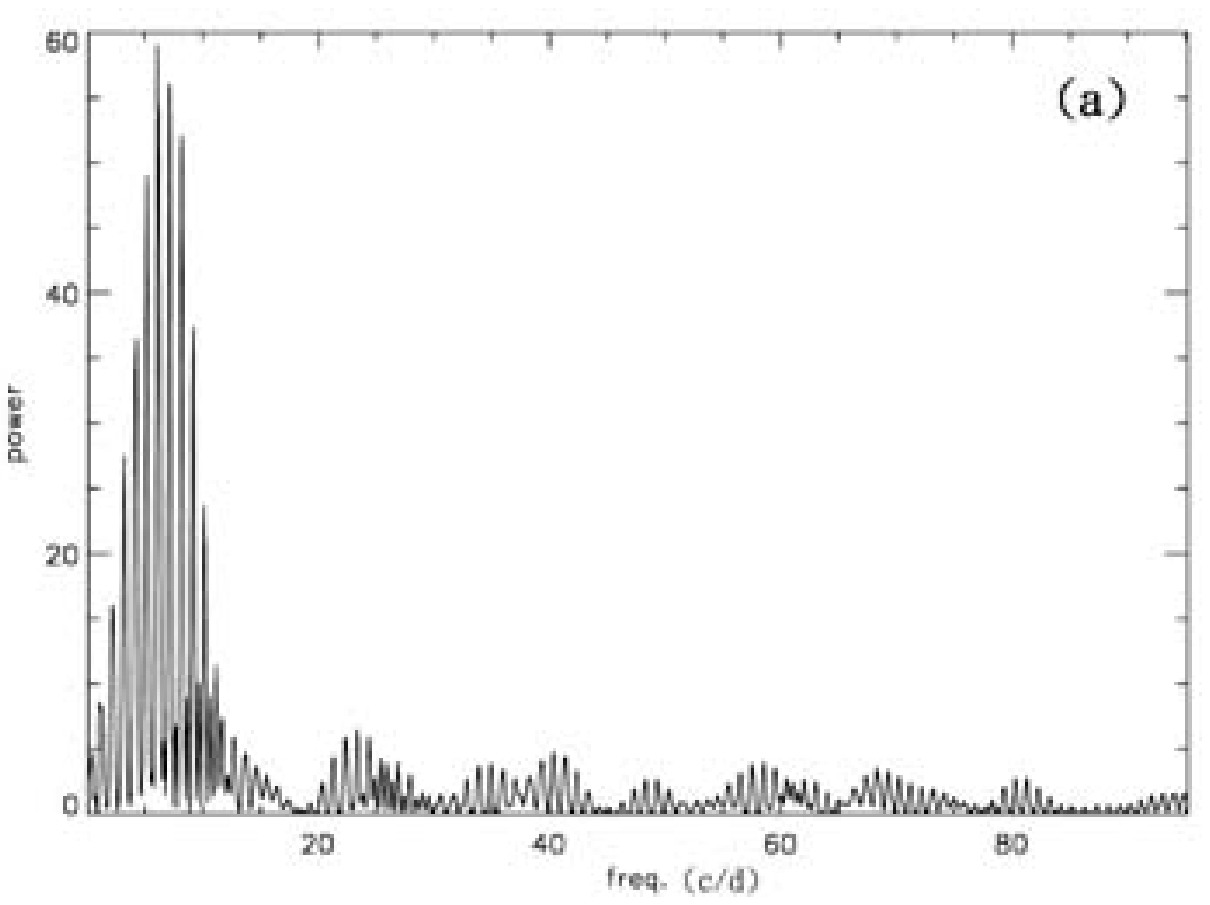,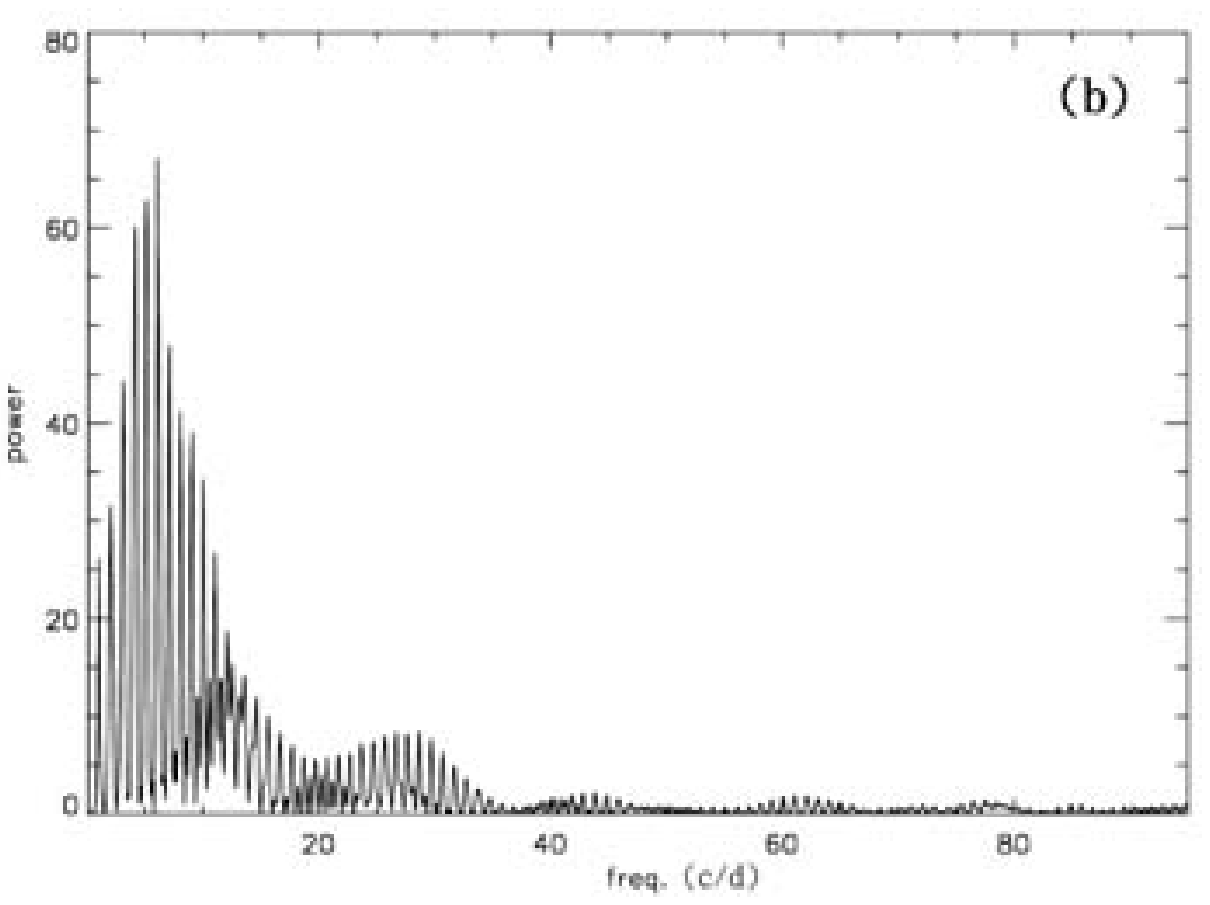]{Power spectra of the time-series photometric data in both R (a) and I (b).
\label{fig.3}}

\figcaption[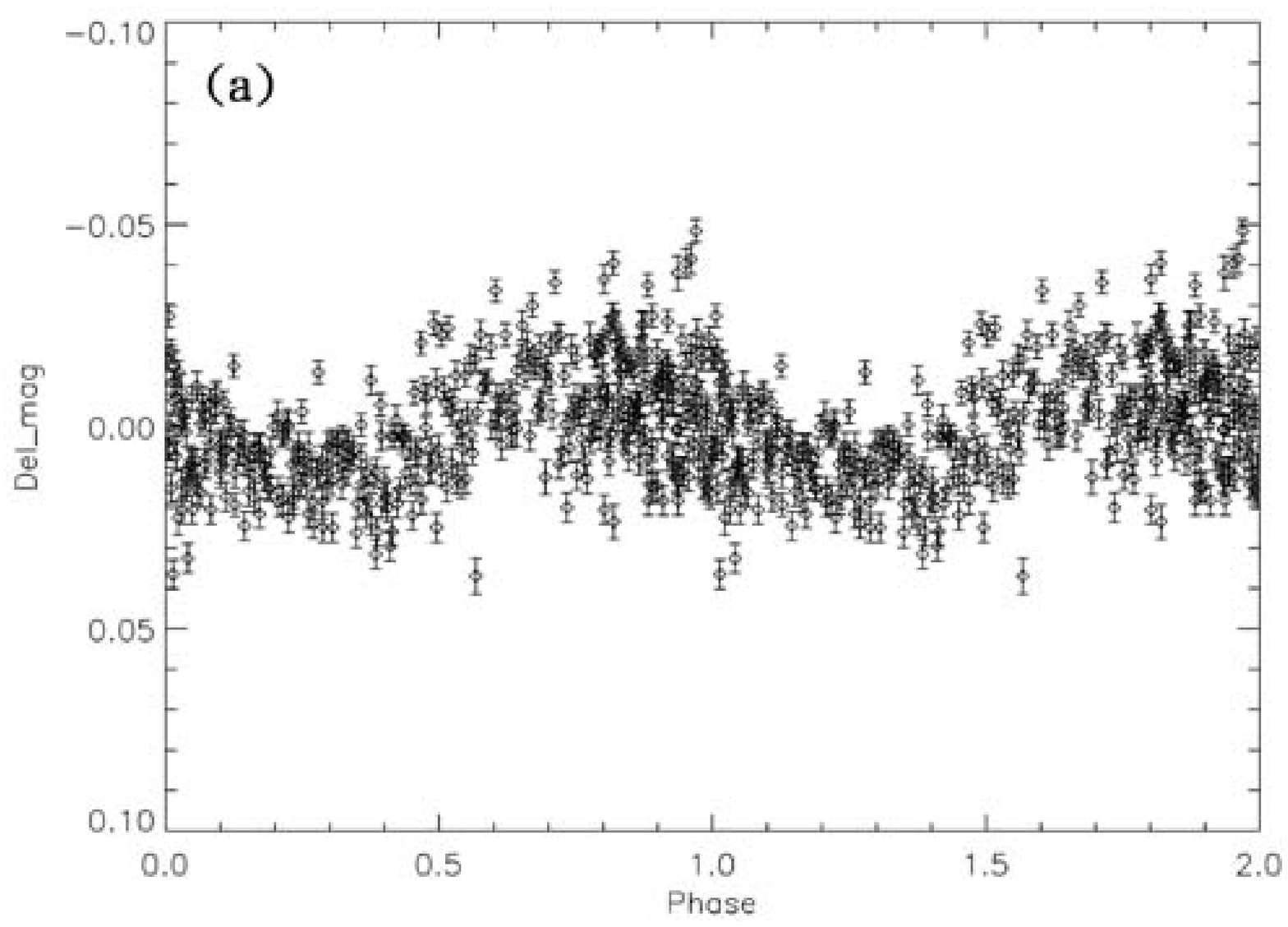,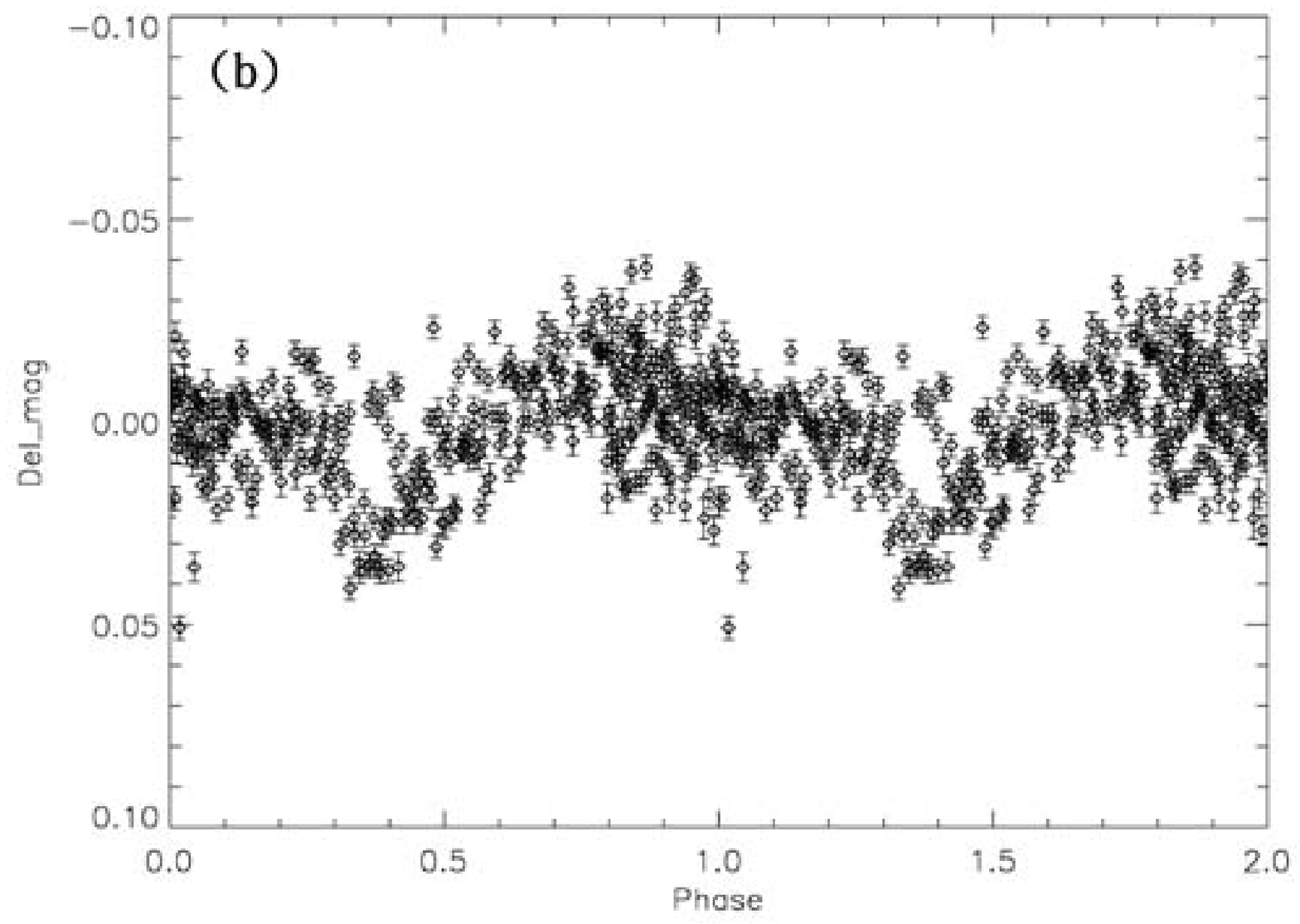]{Phase diagrams of the photometric data in R (a) and I (b).
The periodic variations are believed to be due to an eclipsing binarity nature of the
central source.  The average error of the phase bin is $\pm$0.0074 mag for the R band
and $\pm$0.0046 mag for the I band. \label{fig.4}}

\figcaption[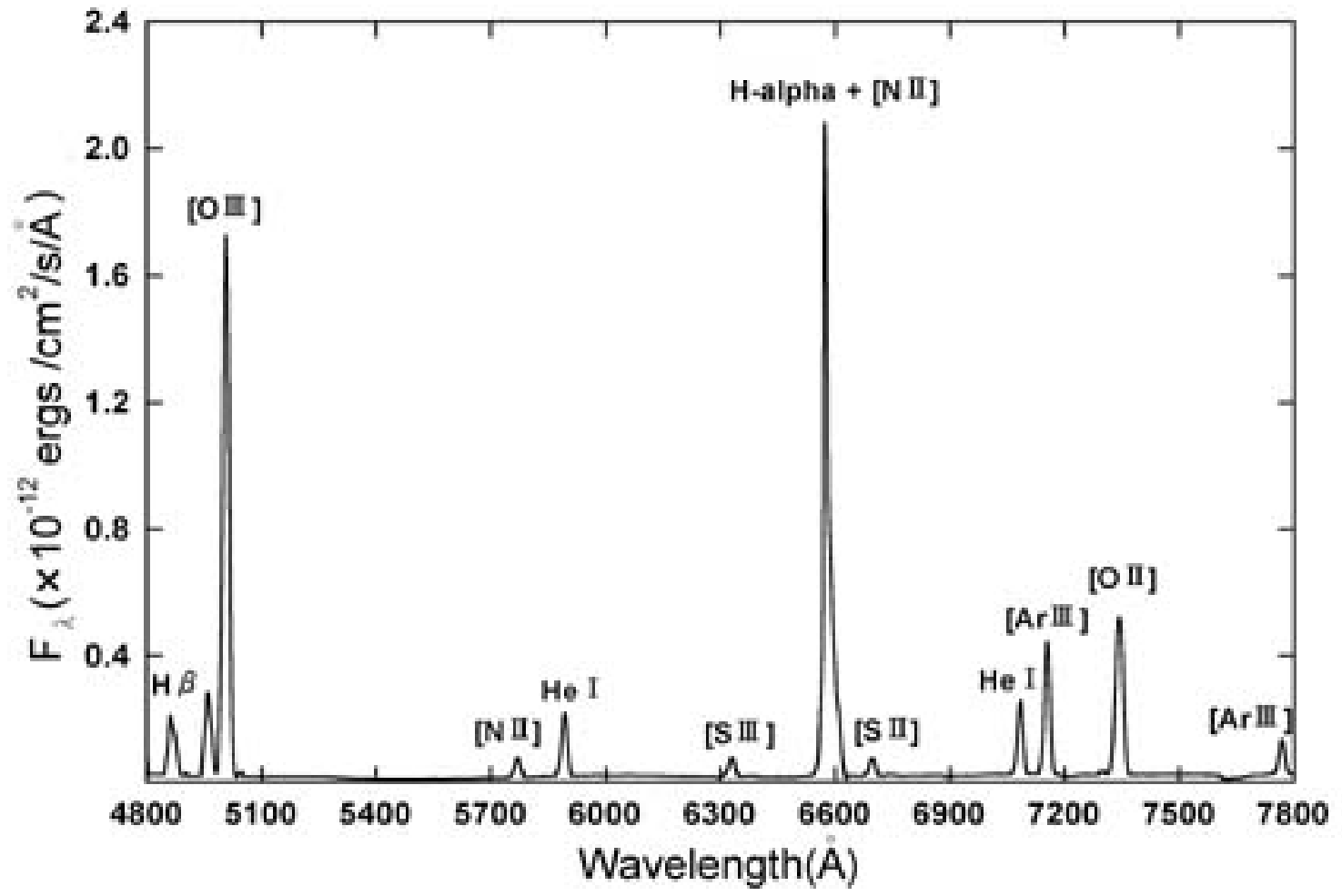]{The spectrum of HB 12 in the wavelength range from 4800 to 7800 \AA.  \label{fig.5}}

\figcaption[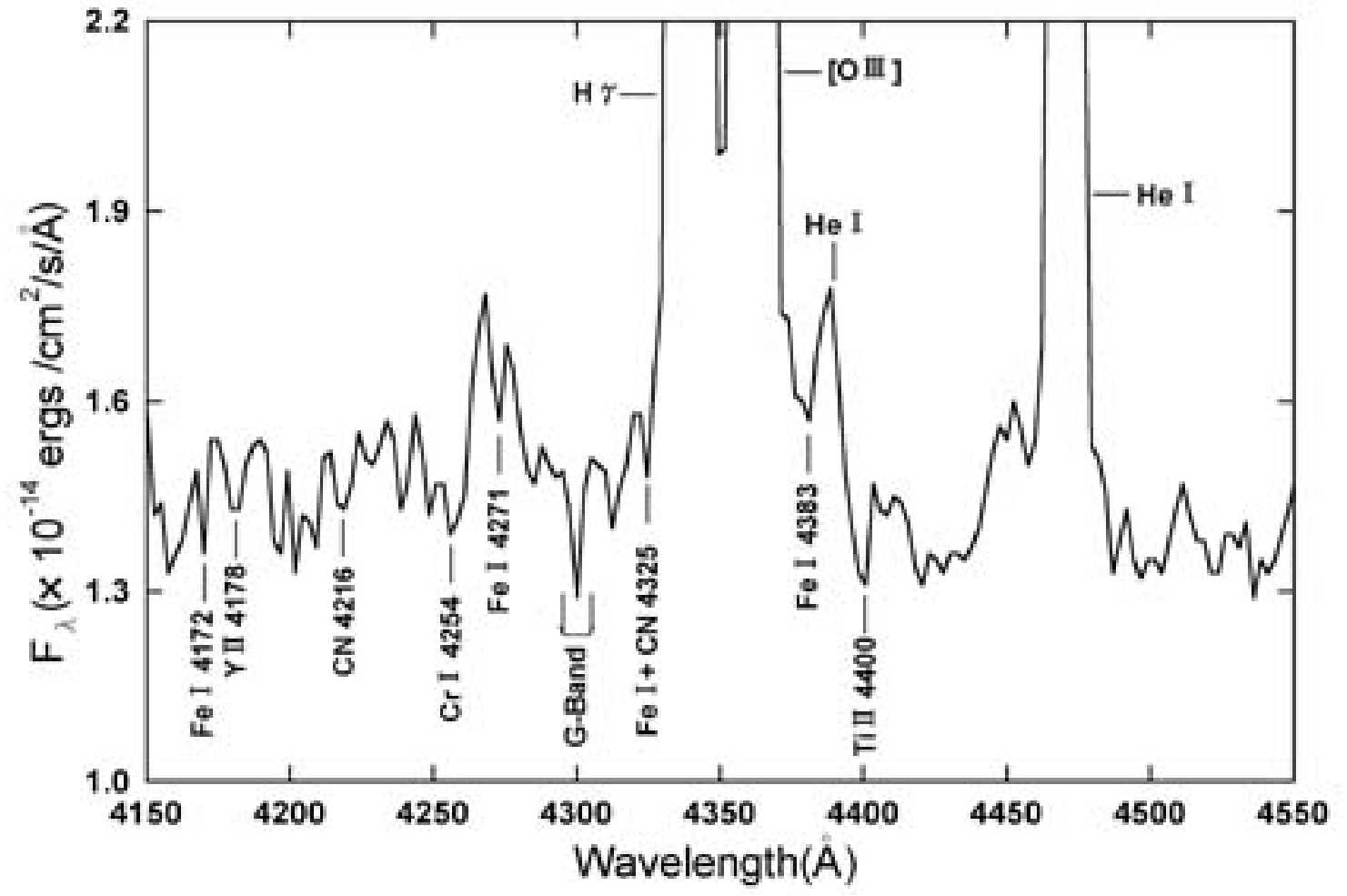,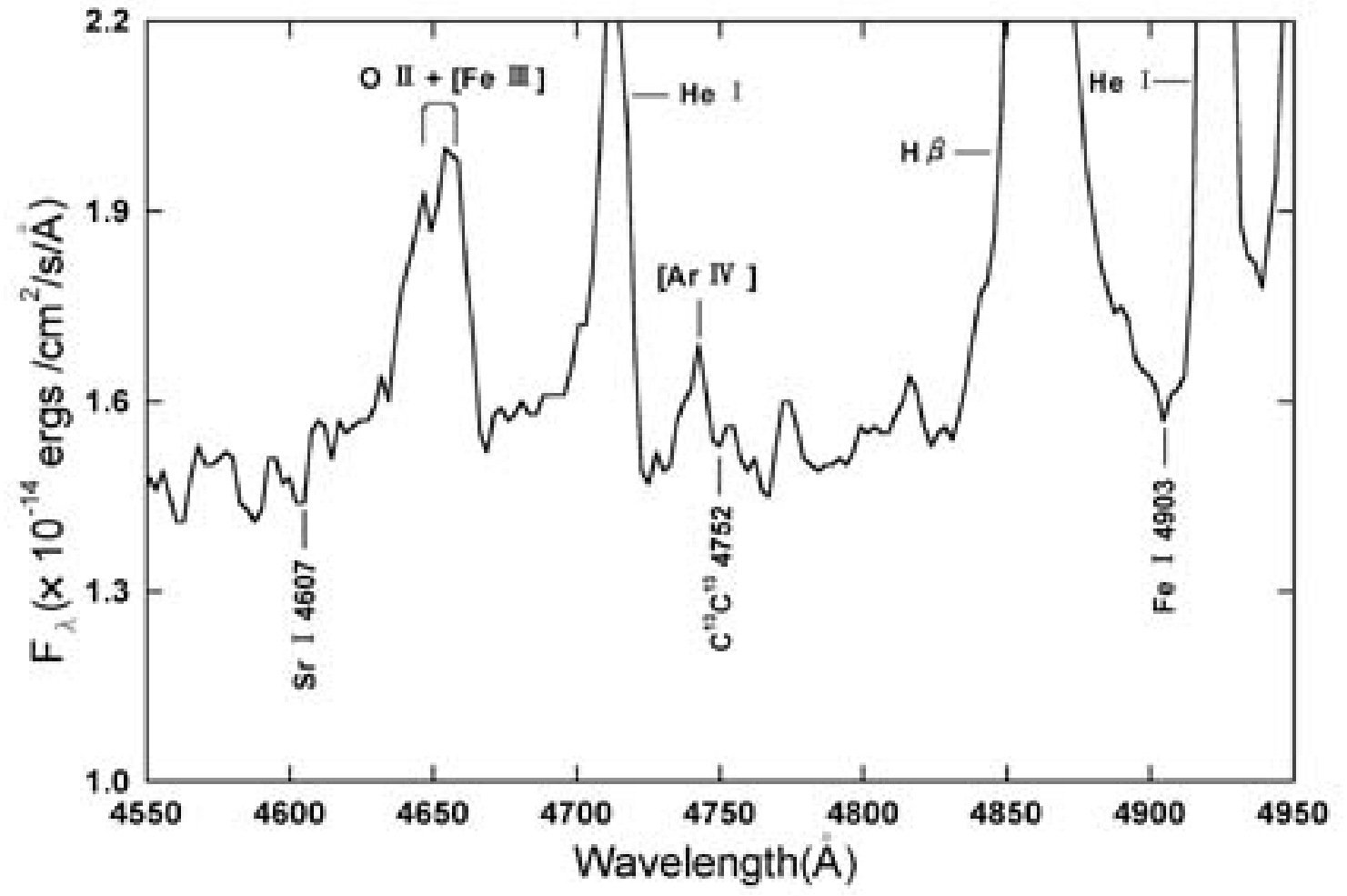,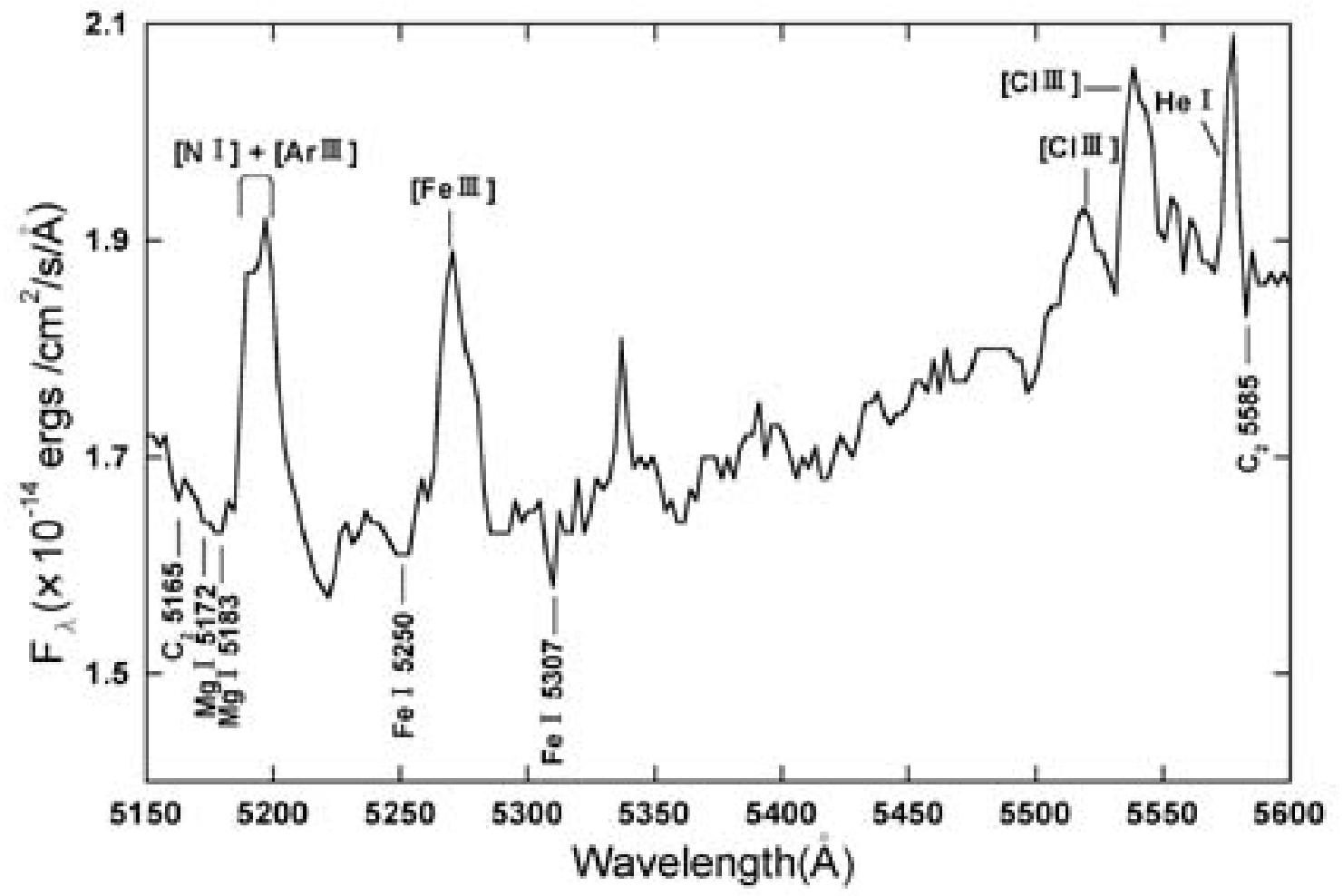]{Enlarged spectrum of HB12 in the ranges (a) 4150 - 4550 \AA,
(b) 4550 - 4950 \AA, (c) 5150 - 5600 \AA. \label{fig.6}}

\newpage
\epsscale{0.85}
\plotone{f1.ps}
fig.1
\newpage
\epsscale{0.925}
\plotone{f2a.ps}
\figurenum{fig.2a}
fig.2a
\plotone{f2b.ps}
\figurenum{fig.2b}
fig.2b
\newpage
\epsscale{1.1}
\plotone{f3a.ps}
\figurenum{fig.3a}
fig.3a
\plotone{f3b.ps}
\figurenum{fig.3b}
fig.3b
\newpage
\epsscale{1.1}
\plotone{f4a.ps}
\figurenum{fig.4a}
fig.4a
\plotone{f4b.ps}
\figurenum{fig.4b}
fig.4b
\newpage
\epsscale{1.1}
\plotone{f5.ps}
fig.5
\newpage
\epsscale{1.1}
\plotone{f6a.ps}
\figurenum{fig.6a}
fig.6a
\plotone{f6b.ps}
\figurenum{fig.6b}
fig.6b
\plotone{f6c.ps}
\figurenum{fig.6c}
fig.6c

\clearpage

\begin{deluxetable}{cccccc}
\tabletypesize{\scriptsize}
\tablewidth{0pt}
\tablecaption{Journal of photometric observations of HB 12}
\startdata
\noalign{\smallskip}
\tableline
\noalign{\smallskip}
Observation date &  Start  &  Start &  Duration  &  Filter & Number of
exposures \\
       &   (UT)   &  (HJD 2452900 +) & (hrs) &    &  \\
\noalign{\smallskip}
\tableline
\noalign{\smallskip}
2003 December 03 & 10:56 & 76.9557 & 4.3 & R & 132 \\
2003 December 03 & 10:57 & 76.9568 & 4.3 & I & 132 \\
2003 December 04 & 10:51 & 77.9527 & 4.4 & R & 139 \\
2003 December 04 & 10:52 & 77.9540 & 4.4 & I & 139 \\
2003 December 05 & 10:38 & 78.9434 & 4.5 & R & 143 \\
2003 December 05 & 10:40 & 78.9452 & 4.5 & I & 143 \\
\noalign{\smallskip}
\enddata
\end{deluxetable}

\begin{deluxetable}{ccccc}
\tabletypesize{\scriptsize} \tablewidth{0pt} \tablecaption{Summary
of Journal spectral observations of HB 12} \startdata
\noalign{\smallskip} \tableline \noalign{\smallskip}
Observation date &  Wavelength  &  Resolution &  Width of Slit & Integration Time \\
       &   (\AA)   &  (\AA~pixel$^{-1}$) & (arcsec) &  (s)  \\
\noalign{\smallskip}
\tableline
\noalign{\smallskip}
2004 Aug 08 & 4800 - 10500 & 3.1 & 3.6 & 300 \\
2004 Aug 08 & 4800 - 10500 & 3.1 & 3.6 & 900  \\
2005 Sep 26 & 3800 - 6200 & 2.4 & 2 & 2 $\times$ 3600  \\
\noalign{\smallskip}
\enddata
\end{deluxetable}

\end{document}